# Electronic Conductivity Upturn of HOPG Contrast to Transport Properties of Polycrystal Graphite

Zhiming Wang[1,2*], Feng Xu[3], Chao Lu[1], He Zhang[1], Qingyu Xu[2], Jinan Zhu[1]

*1. Institute of Mechanical Engineering, Nanjing University of science and technology Nanjing 210094, China*

*2. National Laboratory of Solid State Microstructures and Department of Physics, Nanjing University, Nanjing 210093, China*

*3. Department of Material science and Engineering, Nanjing University of science and technology Nanjing 210094, China*

Abstract:

The transport properties of highly oriented pyrolitic graphite (HOPG) and polycrystal graphite have been studied. The electric conductivity of HOPG is several times larger than that of the polycrystal graphite. Along with the large magnetoresistances (MR), the polycrystal graphite show the accordant semiconductor-like character in a wide temperature (roughly range from 20K to 120K) under 0, 4, 8, 12 T applied magnetic field, while the magnetic-field-induced metal-semiconductor-like transition was only found in HOPG. The difference of transport properties originates from the Coulomb interaction quasipartical in HOPG graphite layers in contrast with the grain boundary scattering in the polycrystal graphite.

* Corresponding author: zhimingwang@mail.njust.edu.cn

Carbon is the most common element in nature. Recently, the discovery of $C_{60}$ and the successful preparation of carbon nanotube have opened a new path for the configuration of Carbon. Both graphite and $C_{60}$ can be electron-doped by alkali metals to become superconducting (transition temperatures up to 52 K). And spontaneous magnetization has occurred in a polymeric $C_{60}$, which is ferromagnetic at room temperatures with higher Curie temperature $T_C$ (up to 500 K)[1], and the magnet may be made from carbon [2]. The discoveries have attracted many interests on carbon-based electronics [3].

Graphite with $sp^2$-bonded configuration is a semimetal, which has a highly anisotropic electronic structure with nearly compensated low-density electrons and holes of very small effective mass. Such an unusual electronic structure is the basis of the unique electronic properties of other graphitic materials, such as Fullerenes and carbon nanotubes, and may lead to novel manifestations in the two-dimensional graphene. The key development in carbon electronics is that these graphite sheets can be rolled up into tubes, which quantizes the momentum of the electrons moving around the circumference of the tube [3]. Many theoretical and experimental works have been done on the electronic transport in graphite [4-5]. In the semimetallic graphite material, there are highly anisotropic Fermi surface, low carrier concentrations, small effective carrier masses, and long carrier mean free path. Because of these properties, a large ordinary magnetoresistance (OMR) effect at room temperature, Shubnikov-de Haas oscillations and the metal-semiconductor-like transition, had been observed in high quality graphite sheet, highly oriented pyrolitic graphite (HOPG)[6]. In this letter, we report an extraordinarily large MR observed in HOPG and the difference of transport property between HOPG and the polycrystal graphite.

The HOPG samples used in this study were supplied from the Advanced Ceramics Corporation. The electron diffraction patterns of HOPG and the polycrystal graphite are shown in Fig. 1 (a), (b), which were taken by transmission electron microscope (TEM, Philips CM12). The figure (a) clearly shows HOPG sample hexagonal structure with a lattice constant amounting to 0.2456 nm. Only (002), (004), (006) peaks were observed in X-ray diffraction patterns of HOPG, while is some other peaks observed in the polycrystal graphite.

The standard four-probe method was used to measure resisitance of HOPG samples with the magnetic fields $H$ up to 8.15 T in the geometry of perpendicular field and the current perpendicular to graphite layer plane (P: H⊥I, H⊥plane), in the geometry of transverse field in the plane but perpendicular to the current (T: H⊥I, H // plane), and in the geometry of longitudinal field parallel to the current (L: H // I, H // plane). The current was applied in the plane of sample layer. The MR is defined as $[\rho(H) - \rho(0)]/\rho(0) \times 100\%$, where $\rho(H)$ is the resistivity in a magnetic field H.

Figure 2 shows the $H$ dependence of resistance in three geometries for HOPG sample and polycrystal graphite at 4.2 K and 300 K, respectively. Several main characteristics can be found: (1) MR ratio is always positive without the hysteresis; (2) MR increases with decreasing temperature; (3) the perpendicular MR is much larger than the transverse and longitudinal MR; (4) the magnetic field dependence of MR is generally quadratic at low field and linear at high field. The perpendicular MR in HOPG sample exhibits extraordinarily huge values amounting to 85,300 % at 4.2 K under 8.15 T and 4,950 % at 300 K, while the perpendicular MR in polycrystal sample is 172 % at 4.2 K under 8.15 T and 117 % at 300 K. These characteristics listed above had indicated the OMR effect in HOPG and polycrystal graphite. The OMR effect exists in all metals, alloys and semiconductor materials due to the curving of conduction carrier trajectories by the Lorentz force under a magnetic field. It is usually quite small in common metals, only in the order of 1 %. To realize a substantial large OMR effect, two necessary conditions are required. The first one is that $\omega_c \tau$ should be at least of order 1, where $\omega_c = eH/m^*c$ is the cyclotron frequency with $m^*$ the effective carrier mass and $\tau$ the relaxation time. Physically, $\omega_c \tau$ indicates the turn angle in radians within the scattering time $\tau$ during the spiraling motion of carriers in a magnetic field. However, the first condition alone cannot guarantee the existence of OMR. In fact, no MR effect is predicted using a free-electron theory with spherical Fermi surface only. Two band structure and/or complex Fermi surface must be required in order to get the OMR, which is another necessary condition [19].

It was noticed from Fig. 2 that the field dependence curve of the MR is not smooth in the perpendicular (P) and possesses a characteristic of the oscillation by comparison to the linear dependence. The oscillation can be attributed to Shubnikov-de Hass effect. The oscillations are periodic with 1/H and could be more clearly observed as the orderliness on Fig. 3 in the reference [6]. Firstly discovered in bulk Bi single crystals in 1930, the Shubnikov-de Hass effect is believed from the diamagnetism and Landau quantization of the cyclotron orbits of the carriers in field. The motion of the conduction carrier turns as trajectories owing to the applied magnetic field, which is the superposition of the linear motion along the field direction and the circular motion in the plane perpendicular to the field. With increasing magnetic field, both the occupancy of each Landau level and the separation between adjacent Landau levels ($\hbar\omega_c$) become larger. As a result, the Landau levels below the Fermi level are sequentially driven across the Fermi level ($E_F$) with a periodical change in the density of states and conductivity. Due to the highly anisotropic Fermi surface of graphite, Shubnikov-de Hass oscillations can be observed clearly in HOPG at the low temperature. With increasing temperature, Fermi surface become illegibility so that the oscillation is from faintness into illegibility, disappearing at last, so the dependence of MR-H observed were linear in the high temperature region [6].

In Fig.3 (a), (b) which the metal-semiconductor-like transitions and semiconductor-like conduction in the HOPG and polycrystal graphite samples were observed in different applied fields over a wide temperature range, respectively. It is different in the transport properties between HOPG and polycrystal graphite. The temperature dependence of resistvity of HOPG in the zero magnetic fields is typical metallic behavior, $d\rho/dT > 0$ at lower temperatures and $d\rho/dT < 0$ at higher ones. The electronic conductivity upturn phenomena in HOPG are more clearly seen for applied field perpendicular to the basal plane, suggesting that the orbital motion of quasiparticales is responsible for the change of the conductivity dependence. In terms of conductivities theory, the component of the resistivity reads [7]

$$\rho_{xx} = \frac{\sigma_{xx}}{\sigma_{xx}^2 + \sigma_{xy}^2} \qquad (1)$$

Where the general behavior of the resistivity were controlled by the value the charge density in two opposite regimes of dynamics, at small density, when the Hall conductivity $\sigma_{xy}$ is negligible compared to $\sigma_{xx}$, the resistivity in Eq. (1) behaves as $1/\sigma_{xx}$; on the other hand, at sufficiently large density, when the Hall conductivity dominates over the component, the resistivity $\rho_{xx} \approx \sigma_{xx}/\sigma_{xy}^2$, thus the electronic conductivity upturn occur and $T_c$ separates a high-temperature semiconducting-like behavior ($\rho_a$ increases with temperature decreasing) from a low-temperature metallic-like behavior ($\rho_a$ decreases with temperature decreasing). Gorbar *et al* gave the value of critical temperature $T_c$ as a phenomenological parameter based on temperature dependence of the resistivity [7].

$$T_c \approx 0.2 v_F \sqrt{|eB|/c} \qquad (2)$$

The magnetic field dependence of the critical temperature suggests that this anomalous increase is not explained by a single-electron picture. Yoshioka and Fukuyama had proposed a model explained this phenomenon in terms of the charge-density-wave (CDW) induced by the Landau quantization in the highly anisotropic energy bands. The electronic band structure in the layered graphite is very anisotropic. With the magnetic field applied, the electron system rapidly gets into the quasiquantum limit, where only a few Laudau levels remain below the Fermi level, and higher Landau levels are far apart. After investigated various types of electronic instabilities, including of CDW, spin-density waves, and an excitonic phase, associated with these four one-dimensional bands by the electron-phonon interaction or the Coulomb interaction, they found that the Coulomb-interaction-induced CDW instability corresponding to the wave number spanning the two Fermi wave vectors of the subband has the higher transition temperature in the magnetic fields [8]. Then the critical temperature $T_c$ is obtained as a function of magnetic field by the YF theory: [9-10]

$$T_c = 4.53 \frac{E_F}{k_B} \frac{\cos^2(c_0 k_{F01}/2)}{\cos(c_0 k_{F01})} \times \exp\left\{-\left[\frac{u}{(2\pi l)^2}\left[\frac{dk_z}{de_{01}}\right]_{E_F}\right]^{-1}\right\} \quad (3)$$

Where $E_F$ is the Fermi energy, $k_{F01}$ is the Fermi wave vector along the $kz$ corresponding to the subband, $c_0$ is the $c$-axis lattice constant, $l$ is the Larmor radius given by $l^2 = \hbar/eB$, $B$ is the magnetic field, and u is the exchange interaction, which is a fuction of the dielectric constant $\varepsilon$. Iye and Berglund *et al.* had investigated the phase diagram in the magnetic field up to 280 kG and temperature down to 100 mK by the magnetotransport measurement, and found that the field dependence of the transition temperature is empirically formula,[10-11]

$$T_c = \alpha \exp(-/\xi B) \quad (4)$$

Where $\alpha$ and $\xi$ are fitting parameters whose values had been determined as 69 K and 1047 kG, respectively. This functional form is basically interpreted a BCS-like expression for $T_c$, where the principal *B*-dependence arises from the *B*-linear increase of the density of state at the Fermi level. Equation (2) is empirically led from the analogy to the BCS-type expression for the transition temperature: [10]

$$T_c = 1.13E_F \exp\left[-\frac{1}{N(E_F)V}\right] \tag{5}$$

Where $N(E_F)$ is the density of states at the Fermi level and V is the pairing interaction, and $\alpha$ corresponds to $E_F$, $\xi$ corresponds to $V$. Since the density of states is almost proportional to the magnetic field, the magnetic field dependence in the exponential factor in Eq. (2) just represents the field dependence of $N(E_F)$. Taking account of the field dependence of $N(E_F)$, Ochimizu et al gave an approximate the preexponential factor $\alpha$ with a linearly decreasing function of field: [8]

$$\alpha(B) = A(19 - 0.25B)K \tag{6}$$

Where A is a fitting parameter, B is in units of Tesla. Thus the transition temperature [8]

$$T_c = A(19 - 0.25B)\exp\left[-\frac{1}{\xi B}\right] \tag{7}$$

Other theoretical analysis suggest that the electronic conductivity upturn in the HOPG is the condensed-matter realization of the magnetic catalysis phenomenon known in relativistic theories of (2+1)-dimensional Dirac fermions. According to the theory, the magnetic field opens an insulating gap in the spectrum of Dirac fermions of graphene, associated with the electron-hole pairing, below a transition temperature which is an increasing function of field. [12-17]

The electronic conductivity upturn behavior is the result of a competition between increased carrier densities and reduced carrier mobility ($d\mu_H/dT < 0$) with temperature increased. In reality, the occurrence of $d\rho/dT > 0$ or $d\rho/dT < 0$ behavior in graphite at the zero magnetic fields is very sensitive to the carrier density.

As a magnetic field is applied on HOPG in the P geometry, the resistivity increases rapidly and exhibits insulating behavior, not only $d\rho/dT < 0$ but $d\mu_H/dT > 0$. The insulating resistivity was found to develop below a well-defined field-dependent temperature, $T(H) = A(H-H_c)^{1/2}$ with $A = 350(K/T^{1/2})$ and $H_c = 0.038$ (in unit of Tesla) being the fitting parameters [18]. For $H > 1$ k Oe, we have $T(H) \approx 350 H^{1/2}$ ( K ), higher than the room temperature, below which the transition from a semimetallic state to an insulating one can be driven by a magnetic field. As a result, the metal-semiconductor-like transitions in HOPG occur along with the extraordinarily large MR effects over a wide range of temperature.

Figure 3 (b) shows the resistivity of polycrystalline graphite sample with the temperature curve in different magnetic field. The trend in R-T curve shows that is semiconductor or insulator type conductivity behavior over a wide temperature range in 0 T and external magnetic fields (4T, 8T, 12T), in which the semimetal-insulator-like transition do not occur. As quasi-two-dimensional (2D) semimetal with the low carrier density, HOPG can be considered as strong correlation system so that Coulomb interaction can not be ignored transport properties. Then the magnetic field induced behavior plays an important role in HOPG transport property at the low temperature. However the Coulomb interaction is not very evident in the high temperature, then insulating transport properties was dominant in the high temperature, then the electronic conductivity upturn i.e. metal-semiconductor-like transition occur in HOPG. The grain boundary scattering exist in polycrystalline graphite samples, and the Coulomb effect is not so obvious, so there was not clear metal-semiconductor-like transition in polycrystalline graphite samples.

Acknowledgement:

The work was supported by National Laboratory of Solid State Microstructure of Nanjing University (LSSMS) under Grant No.M041901, No.M06004, and Department of personnel Jiangsu Six Talent Fund under Grant No. AD 41118.

Figure Caption:

Fig.1 a TEM diffraction images of HOPG b TEM diffraction images of Polycrystal.

Fig. 2 MR of HOPG、polycrystal graphite vs. the applied field measured at 4.2K and 300 K( P: perpendicular H⊥I ,H⊥plane    T: transverse H⊥I ,H∥plane    L: longitudinal H∥I，H∥plane)

Fig. 3 a Basel-plane resistance measured in the polycrystal graphite sample in the applied magnetic field 0、4、8、12 Tesla regimes, respectively. b Basel-plane resistance measured of HOPG in the applied magnetic field 0、4、8、12 Tesla regimes, respectively.

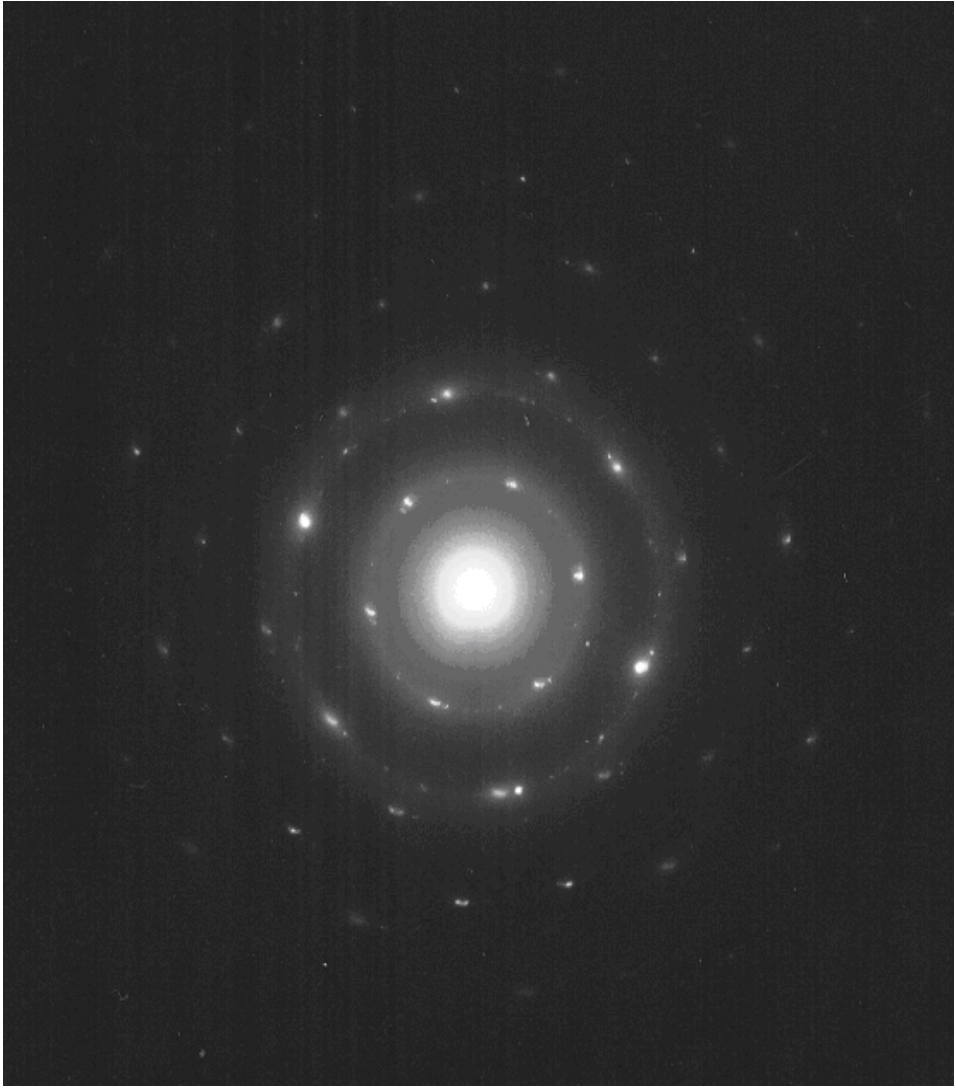

**Figure**

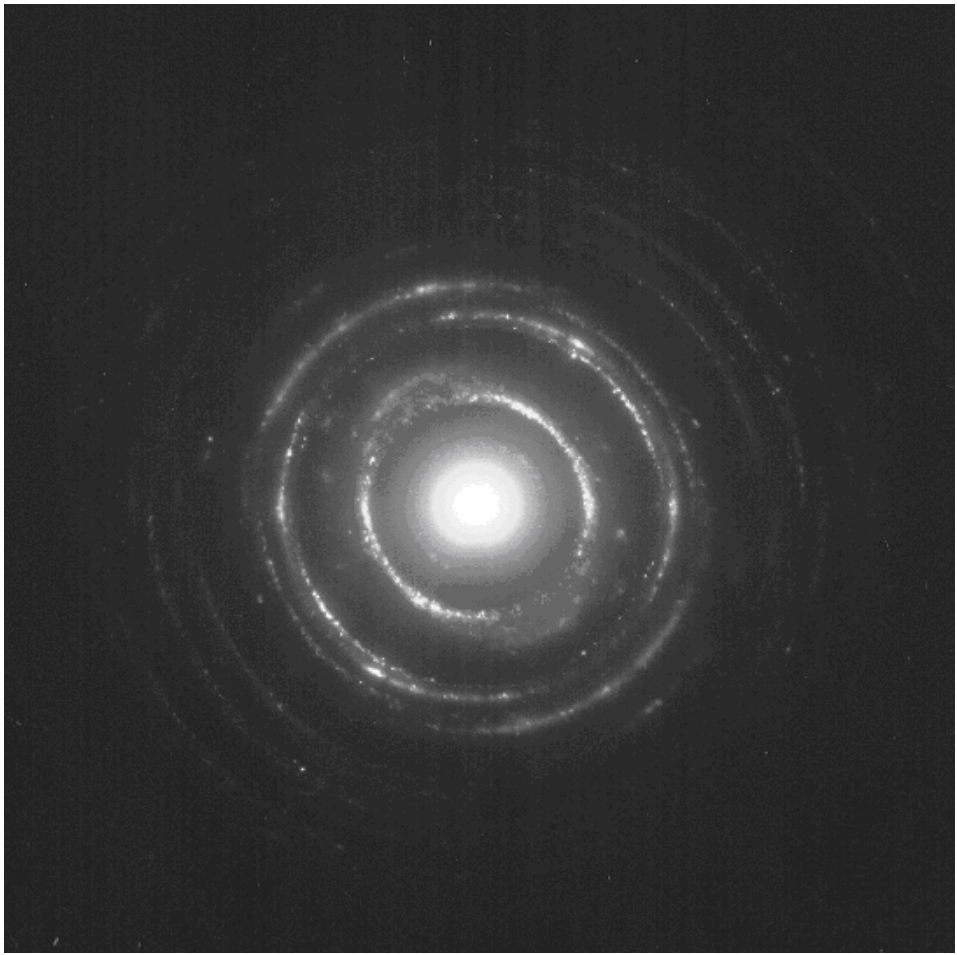

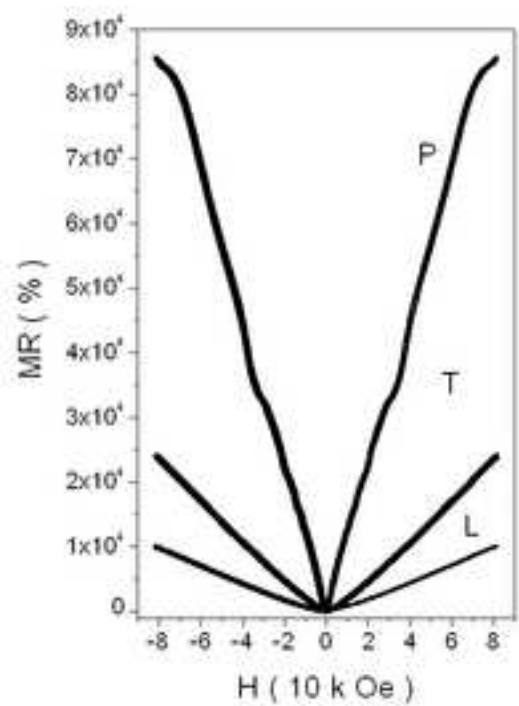
a  HOPG 4.2 K

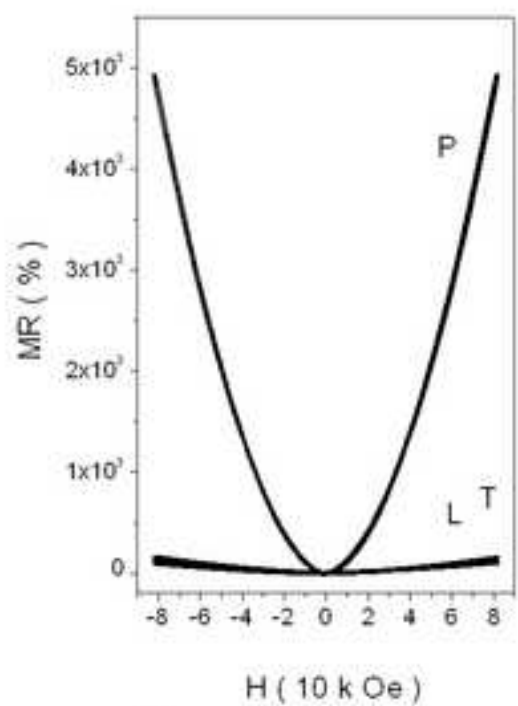
b  HOPG 300 K

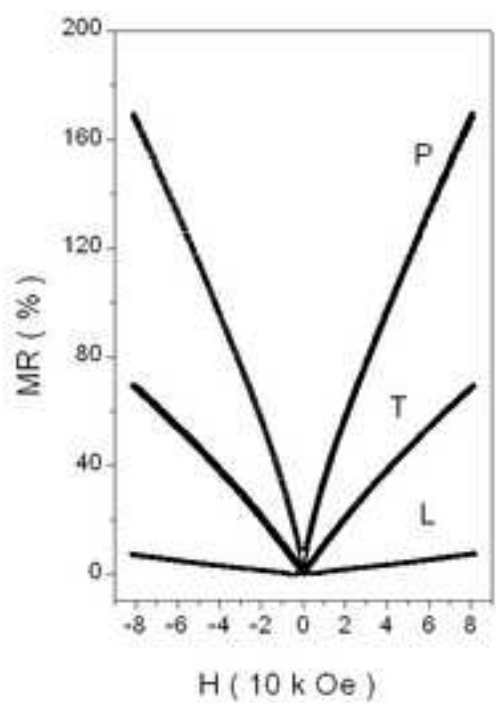
c  polycrystal  4.2 K

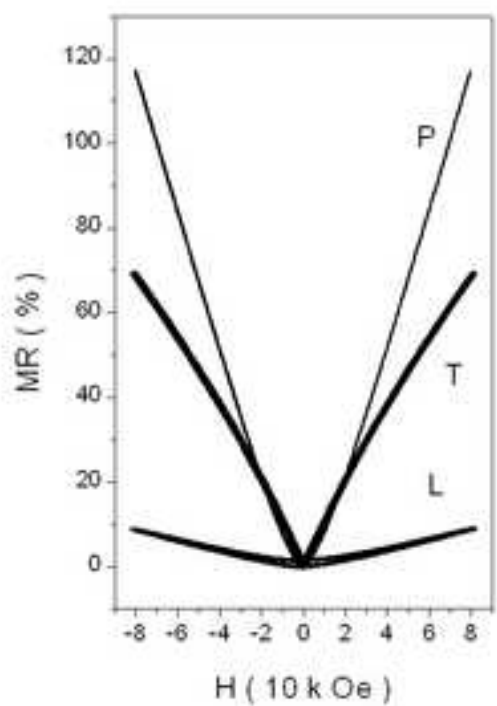
d  polycrystal  300 K

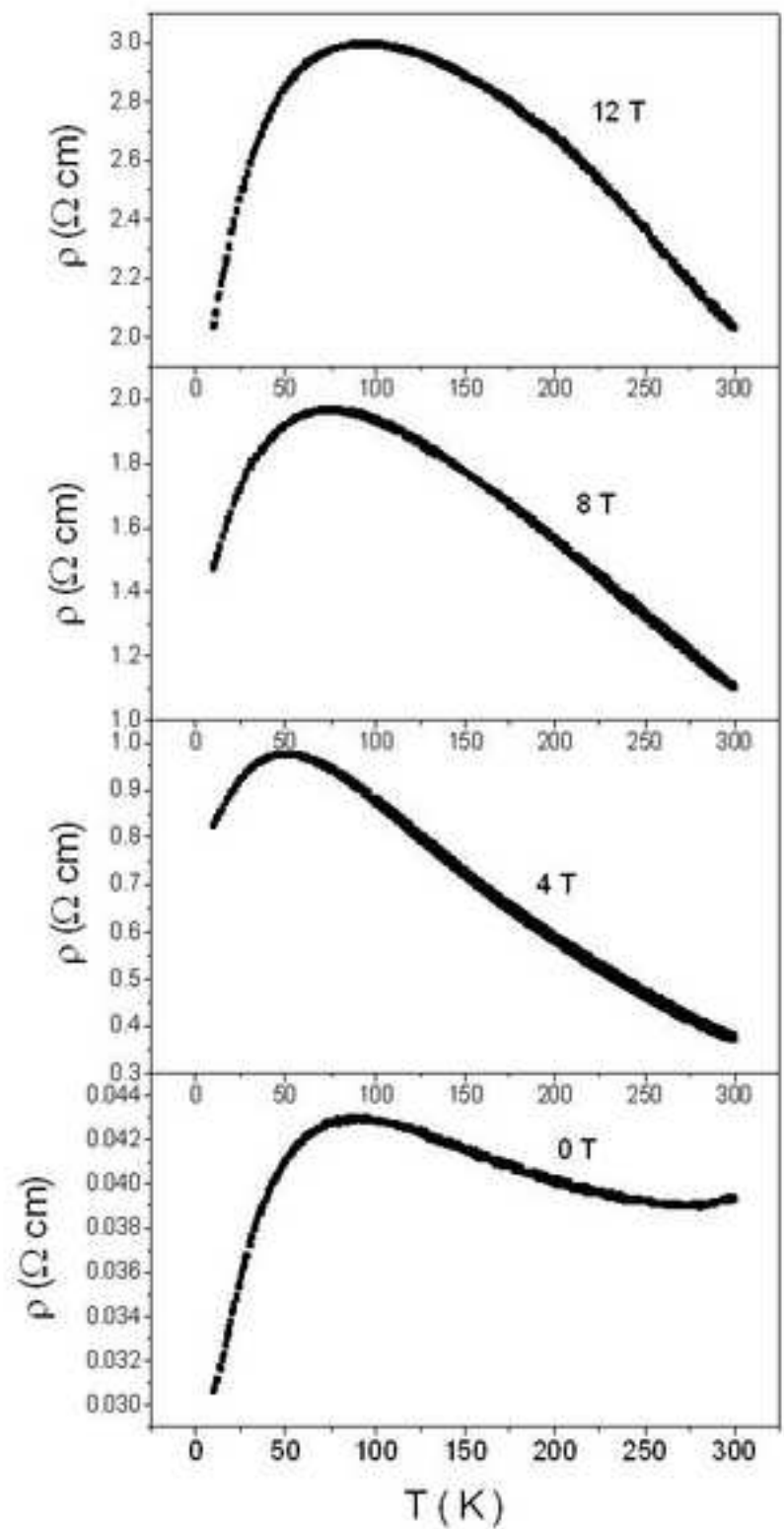

Fig. 3 a Zhiming Wang et al

**Figure**
**Click here to download high resolution image**

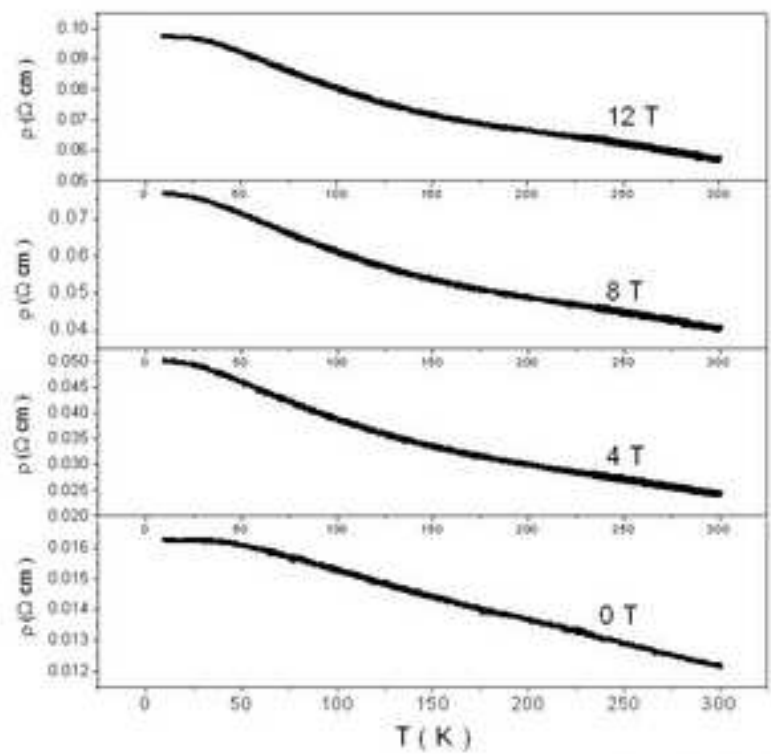

Fig. 3 b Zhiming Wang et al